\documentclass[aps,prl,twocolumn,groupedaddress,showpacs]{revtex4}
\usepackage[dvips]{graphicx}
\usepackage{amsfonts}

\begin{document}

\title{Orthogonality Catastrophe for Vortices in $d$-Wave Superconductors}

\author{Akakii Melikidze}
\affiliation{Kavli Institute for Theoretical Physics, University of California,
        Santa Barbara CA 93106, USA}

\date{December 29, 2002}

\begin{abstract}
The dynamics of magnetic vortices in the mixed state of $d$-wave
superconductors is affected by interaction with quasiparticles near
the gap nodes. We study this effect by computing the overlap of the
ground state wave functions of the nodal quasiparticles in a
two-dimensional $d$-wave superconductor as the ground state changes in
response to the motion of the vortex. We find that the overlap is
strongly suppressed. This orthogonality catastrophe is specific to a
half-quantum vortex and disappears in the case of a doubly-quantized
vortex. This implies strong inhibition of the motion of half-quantum
vortices. The results allow us to develop a simple description of the
quantum vortex creep at low temperatures.
\end{abstract}


\maketitle



Magnetic vortices in Cuprate superconductors~\cite{Bla,YesMalSha} behave
rather differently compared to their counterparts in conventional
materials. One of the characteristic features of the superconducting
state of the Cuprates is the existence of the low-energy
quasiparticles near the nodes of the $d$-wave gap. One is naturally
led to ask: what is the influence of the interaction between the
vortices and nodal quasiparticles on vortex dynamics?


Before answering this question, let us briefly describe our current
knowledge. Vortex creep at low temperatures is believed to proceed by
quantum tunneling. It has been argued that two forces govern vortex
dynamics during tunneling: the Magnus force~\cite{AoTho} and the
friction force. Their ratio determines the ``cleanness'' of the
material. The creep in the dirty,~\cite{BlaGesVin}
super-clean,~\cite{Fei} and intermediate
regimes~\cite{Creep_intermediate} has been studied theoretically.

Quasiparticles play prominent role in the vortex dynamics. In $s$-wave
superconductors, finite gap in the bulk precludes the existence of
low-energy extended states, however quasiparticle states localized in
the cores of vortices, where the gap vanishes,
exist.~\cite{Core_States} Transitions between these states give rise
to the friction force and the renormalization of the Magnus
force.~\cite{KopKra} The situation with quasiparticle states localized
inside vortex cores in $d$-wave superconductors is far less
clear.~\cite{Franz_Tesanovic} On the other hand, extended states exist
and are expected to strongly interact with vortices. Indeed, they have
been argued~\cite{Ashvin,Vafek} to give a quantized value of thermal
Hall conductance in the mixed state at low temperatures. This
prediction is hard to verify since current experiments~\cite{Zhang}
are performed at relatively high temperatures.~\cite{Durst} Quantum
vortex creep, which is also expected to be sensitive to the
interaction with quasiparticles, occurs at much lower temperatures and
thus offers experimental information about the quasiparticle spectrum
at a much lower energy scale.

To study the contribution of nodal quasiparticles to vortex action we
calculate the overlap of the ground state wave functions of the nodal
fermions as the ground state changes in response to the motion of the
vortex. We find that this overlap is extremely small in the case of a
half-quantum vortex. This implies that the motion of half-quantum
vortices is strongly suppressed due to the interaction with nodal
quasiparticles. Indeed, we argue that this orthogonality catastrophe
provides the dominant contribution to the vortex action at
temperatures relevant to quantum vortex creep. Based on this, we
present a simple phenomenological description of the available
experimental data.

Let us consider a vortex in a 2D $d$-wave superconductor. We use the
Bogoliubov-de~Gennes (BdG) equation to treat the interaction of
quasiparticles with the vortex.~\cite{Ashvin,Mel'nikov}. We first take
advantage of the conservation of spin in the BCS theory by changing
from the original electronic operators $c$ to the new fermions:
$\psi_{1k}=c_{k\uparrow}$, $\psi_{2k}=c_{-k\downarrow}^\dag$.  The new
operators are combined to form a two-component spinor: $\psi =
(\psi_1,\psi_2)$. The density of the $\psi$-particles $\psi^\dag\psi$
is the $z$-component of the spin density of the original electrons,
and visa versa. Next, we concentrate on the variation of the phase of
the order parameter $\phi(\vec r)$ around the vortex and neglect the
variation of the amplitude. We then eliminate the phase winding by
making a {\it single-valued} gauge transformation:
$\psi_1\to\exp[-i\phi(\vec r)]\psi_1$, $\psi_2\to\psi_2$. The BdG
equation takes the form $H\psi = \varepsilon\psi$ with the following
Hamiltonian (we use units with $\hbar=c=1$):
\begin{equation}
\label{Hamiltonian}
H = E(\vec p + \vec a +\vec j_s\sigma_z)\sigma_z
    + \Delta(\vec p + \vec a)\sigma_x.
\end{equation}
Here, $\sigma_z$, $\sigma_x$ are Pauli matrices, $\vec p=(p_x,p_y)$ is
the 2D momentum, $E(\vec p)$ is the bare electron energy
relative to the Fermi energy and $\Delta(\vec p)$ is the gap
function. The presence of the vortex produces two contributions to the
Hamiltonian: 1)~There appears a ``Doppler shift''. The superflow
$\vec j_s=\vec\partial\phi/2-e\vec A$ shifts the energies of the nodal
quasiparticles similar to the classical Doppler effect. 2)~The other
contribution is of topological nature. It requires that a parallel
transport of a quasiparticle around the vortex should lead to the
change of its phase by $\pi\nu$, where $\nu$ is the vortex quantum
number.  This is enforced by the Chern-Simons gauge field $\vec a =
\vec\partial\phi/2$ which satisfies $\vec\partial\times\vec a =
\pi\nu\hbar\delta(\vec r)$ (assuming the vortex is at the origin).


First, we would like to calculate the overlap between the ground
state wave functions of the superconductor with and without the
half-quantum vortex. The two terms in the Hamiltonian that were
discussed above both perturb the system shifting the single
quasiparticle eigenstates. An important difference between them is
their spatial range. The superflow $\vec j_s(\vec r)$ is screened at
distances larger than the penetration depth. On the contrary, the
field $a(\vec r)$ decays as $1/r$ at all distances. Let us consider
the Doppler shift term first.  The solution of the orthogonality
catastrophe problem~\cite{OC} in the case of a short-range perturbing
potential is given by:
\begin{eqnarray}
\label{Overlap}
|\langle f|i\rangle| & \sim & N^{-A},\\
\label{Exponent}
A & = & \frac 12 {\rm Tr}
      [\hat\delta(\varepsilon_F)/\pi]^2.
\end{eqnarray}
Here, $|i\rangle$ and $|f\rangle$ are the ground states of the system
before and after the perturbing potential is switched on, $N$ is the
total number of fermions in the system, and
$\hat\delta(\varepsilon)=(1/2i)\ln\hat S(\varepsilon)$ is the phase
shift matrix defined as the logarithm of the scattering matrix.  The
trace in the Eq.~(\ref{Exponent}) is over all states at the Fermi
energy. In our case the Fermi energy is zero. The density of states of
nodal quasiparticles vanishes linearly with energy: there are no
states at zero energy. This property of the phase space implies that
the orthogonality exponent $A$ is zero for any scattering potential of
finite range. This means that the Doppler shift term causes possibly
quite large but finite reduction of the overlap.
Absence of infrared divergence in this case has been found
previously~\cite{Cassanello_Fradkin} in a different context.


Based on what was said above, we set in the following $\vec j_s=0$ in
Eq.~(\ref{Hamiltonian}).~\cite{Mel'nikov} The remaining perturbation
of the system is the long-range vector Chern-Simons field $\vec a(\vec
r)$. Because the field decays slowly at infinity, it is not clear
whether the above result for the orthogonality catastrophe for
short-range potentials can be generalized to this case. It is also not
clear whether at all one can use plane waves as asymptotic states to
calculate the $\hat S$ matrix: the asymptotic completeness has been
proven only for short-range potentials in the Dirac
equation.~\cite{Thaller}

For these reasons we shall adopt another approach.
The field $\vec a(\vec r)$ can be gauged away by a gauge transformation:
$\psi \to \psi \exp(i\nu\theta/2)$,
where $\nu$ is the vortex quantum number and $\theta$ is the polar
angle. The BdG Hamiltonian Eq.~(\ref{Hamiltonian}) reduces to a free one:
\begin{eqnarray}
\label{Free}
H = E(\vec p)\sigma_z+\Delta(\vec p)\sigma_x.
\end{eqnarray}
For a doubly-quantized vortex, $\nu=2$, this implies no orthogonality
catastrophe between the ground states with and without such a
vortex. On the other hand, for a singly-quantized vortex this
transformation makes $\psi$ non-single valued. To take this into
account, we introduce a cut ${\cal C}$ in the $(x,y)$ plane as shown
in Fig.~1.  The vortex is assumed to be at point $A$ at the
origin. Since the cut should end on another half-quantum vortex, we
introduce one at point $B$ for definiteness. We then impose a
$\pi$-phase shift across the cut.

\begin{figure}
\label{Cut}
\includegraphics[width=150pt]{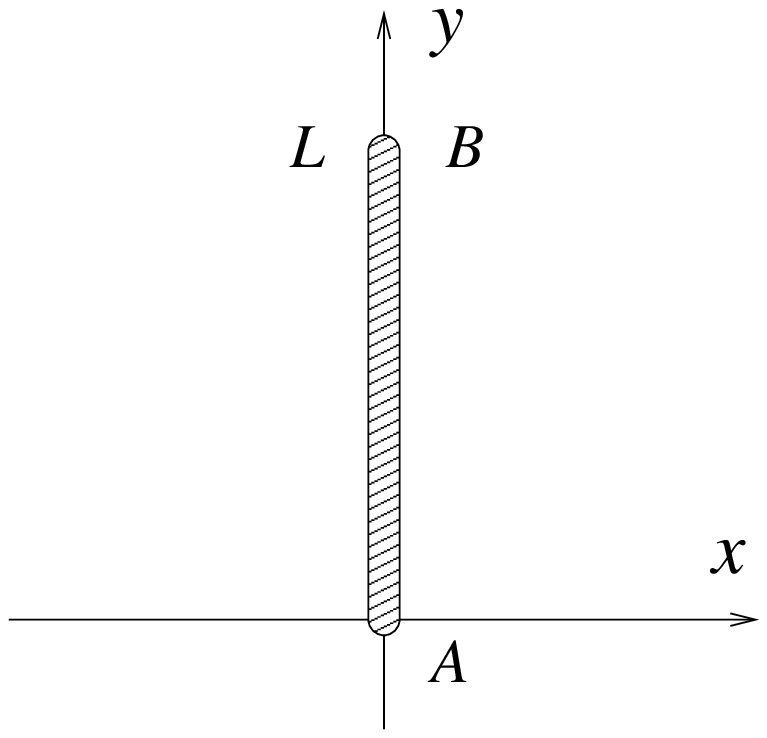}
\caption{The cut ${\cal C}$.}
\end{figure}

The distance $L$ between the two vortices can be interpreted in
various ways: First, $L$ may be thought of as a long-distance
cut-off. In another setting, one may consider an isolated vortex being
moved from point $A$ to point $B$. This can be done by removing a
vortex at point $A$ (inserting a vortex with negative vorticity) and
creating one at point $B$. In this situation $L$ is the distance by
which the vortex has been moved.

Having made the gauge transformation we can reformulate our original
problem as that of finding the overlap of the ground state wave
functions of the system before and after a $\pi$-phase shift across
the cut shown in Fig.~1 was imposed. To this end, consider the {\it
dynamical overlap} function which is defined as follows: Let
$|i\rangle$ be the ground state wave function of the unperturbed
system. We shall work in imaginary time to simplify calculations and
avoid problems with time-ordering.  At time $\tau = 0$ the phase shift
$\pi$ across the cut $\cal C$ is imposed.  Let $|\tau\rangle =
U(\tau)|i\rangle$, where $U(\tau)$ is the imaginary time evolution
operator. Since in the limit $\tau\to\infty$ only the contribution
from the new ground state to $U(\tau)$ survives, we have:
$\lim_{\tau\to\infty}\langle\tau|i\rangle = \langle f|i\rangle$, where
$|f\rangle$ is the new ground state. Since $\langle\tau|i\rangle =
\langle i| U^\dag(\tau)|i\rangle$, the problem reduces to that of
finding the expectation value of the evolution operator in the old
ground state. We have:
\begin{eqnarray}
\label{U_of_tau}
U(\tau) &=& \exp\left\{ i \pi Q(\tau) \right\},\\
Q(\tau) &=& \int_0^\tau J(\tau')\, d\tau'.
\end{eqnarray}
Here, $J(\tau)$ is the total current through the cut $\cal C$ and
$Q(\tau)$ is the net charge that has flowed through the cut between
times $0$ and $\tau$ (here and below, unless mentioned otherwise, the
charges and currents are those of the $\psi$-fermions). The meaning of
the above expressions is quite transparent: Each fermion contributes
phase $\pi$ to the path integral every time it crosses the cut. The
operator $U(\tau)$ can thus be thought as a generator of counting
statistics.~\cite{Levitov_Lee_Lesovik}

Since the unperturbed Hamiltonian is quadratic
in the fermionic currents we can apply Gaussian averaging to obtain:
\begin{eqnarray}
\label{Gaussian}
\langle\tau|i\rangle = \exp\left\{ - A \right\},\\
\label{Gaussian_action}
A = \frac{\pi^2}{2} \langle Q^2(\tau)\rangle.
\end{eqnarray}
To proceed, we need to find the current-current correlation
function. We linearize the Hamiltonian Eq.~(\ref{Free}) around one of
the nodes (e.g. one at $(k_F,0)$): $E(\vec p) = v_F p_x$, $\Delta(\vec
p) = v_\Delta p_y$.  The anisotropy parameter $\alpha =
v_F/v_\Delta\sim E_F/\Delta$ is quite large ($\alpha\approx 14$ for
YBCO). We then make $\alpha=1$ by rescaling the distances: $p_x\to
p_x/v_F$, $p_y\to p_y/v_\Delta$.  This allows us to exploit the full
``relativistic'' invariance of the problem. Indeed, the imaginary time
action (in the second-quantized notation) becomes:
\begin{eqnarray}
\label{Action}
S = \sum\limits_{p} \bar\psi\,\sigma_\mu p_\mu \psi.
\end{eqnarray}
Here, $p_\mu = (w,p_x,p_y)$ is the 2+1 momentum, $\bar\psi =
-i\psi^\dag\sigma_y$. Pauli matrices $\sigma_\mu$ form a Clifford
algebra: $\{\sigma_\mu,\sigma_\nu\}=2\delta_{\mu\nu}$.  The
``relativistic'' invariance is now explicit since both $\sigma_\mu$
and $p_\mu$ transform as 2+1 vectors under rotations of the Euclidean
space-time and the action $S$ is therefore a scalar.  The conserved
current is:
\begin{eqnarray}
\label{Current}
j_\mu = \bar\psi\sigma_\mu\psi.
\end{eqnarray}
The current-current correlator is given (in the free theory) by the
so-called bubble or vacuum polarization diagram:
\begin{eqnarray}
\label{Correlator}
\langle j_\mu j_\nu \rangle_k
 = \int\frac{d^3q}{(2\pi)^3} {\rm Tr} \left[
   \sigma_\mu G(q)\sigma_\nu G(q+k)\right].
\end{eqnarray}
Here, $G(q) = \sigma_\mu q_\mu/q^2$ is the Green function of the Dirac
equation. The correlator, as written, is given by an ultra-violet
divergent expression. This divergence can be eliminated by an
appropriate high-energy cut-off. Introduction of such a cut-off,
however, will break the gauge invariance. A more physical way is to
explicitly impose the current conservation condition: $k_\mu\langle
j_\mu j_\nu\rangle_k = 0$. This way one obtains the well-known
result (see~\cite{QED_3} for a related calculation):
\begin{eqnarray}
\label{Correlator_result}
\langle j_\mu j_\nu \rangle_k
 = \frac{1}{16 k} \left(k^2\delta_{\mu\nu} - k_\mu k_\nu\right).
\end{eqnarray}
The inverse Fourier transform of this correlator is:
\begin{eqnarray}
\label{Inverse}
\langle j_\mu(r) j_\nu(0) \rangle
 = - (\partial^2\delta_{\mu\nu} - \partial_\mu\partial_\nu)
     \frac{1}{32\pi^2r^2},
\end{eqnarray}
where $r_\mu=(\tau',x,y)$. The fluctuation of the charge
$\langle Q^2(\tau)\rangle$ that flowed
through the cut $\cal C$ of length $L$ during time $\tau$
is then given by the integral over the
segment $\Omega=\{ r: 0<\tau'<\tau,\, x=0,\,0<y<L\}$ of the
hyperplane $(\tau,y)$. Explicitly, one has:
\begin{eqnarray}
\label{Fluctuation}
\langle Q^2(\tau)\rangle
 = \int_\Omega dr_1 \int_\Omega dr_2\,
   \langle j_x(r_1) j_x(r_2) \rangle,
\end{eqnarray}
Substituting Eq.~(\ref{Inverse}) into Eq.~(\ref{Fluctuation}) and using
Gauss theorem twice, one obtains:
\begin{eqnarray}
\label{Gauss}
\langle Q^2(\tau)\rangle
 = \frac{1}{32\pi^2} \int_{\partial\Omega}\int_{\partial\Omega}
   \frac{d S(r_1) d S(r_2)}{(r_1-r_2)^2}.
\end{eqnarray}
Here, $dS = \wedge dl$ is the ``surface'' element of the boundary
$\partial\Omega$, while $dl$ is the length element of the
corresponding perimeter. The integral in Eq.~(\ref{Gauss}) diverges as
$r_1\to r_2$. This means that the exponent in the final result
acquires an explicit cut-off dependence. A suitable way to
introduce a short time cut-off is to replace:
\begin{eqnarray}
\label{Replace}
\frac{1}{(r_1-r_2)^2} \to \frac{1}{(r_1-r_2)^2+a^2}.
\end{eqnarray}
The cut-off $a$ is, roughly, the inverse of the energy at which the
linear Dirac dispersion breaks down. It would be justified to assume
$a \sim 1/\Delta$.  The integral in Eq.~(\ref{Gauss}) can now be
easily evaluated. Plugging the result in Eq.~(\ref{Gaussian_action})
in the case where $L,\,\tau\gg \Delta^{-1}$ one obtains:
\begin{eqnarray}
\label{Result}
A \sim \Delta\left(L +\tau \right).
\end{eqnarray}
This can be called a perimeter law for vortex action: the contribution
of nodal fermions to the imaginary time action of a half-quantum
vortex is proportional to the length of the vortex trajectory in the
Euclidean space-time. The proportionality of the action to the
perimeter of $\Omega$ rather than its surface area is essentially due
to the conservation of the number of $\psi$-particles. In terms of the
original electrons, it is the conservation of the $z$-component of
spin polarization.



Let us now discuss the implication of these results for the vortex
dynamics in $d$-wave superconductors. First of all, Eq.~(\ref{Result})
shows that the orthogonality catastrophe for a half-quantum vortex is
very strong. Even when the length $L$ is finite (e.g. when the vortex
moves over a small distance) the dynamical overlap of the two states
vanishes in the limit $\tau\to\infty$. Certainly, finite temperature
provides a cut-off: $\tau_{th}\sim 1/T$. This leads to the conclusion
that the contribution of the nodal fermions to the imaginary time
action at low temperatures is dominated by the second term in
Eq.~(\ref{Result}): $A \sim \Delta/T$.  This, in turn, means that the
dynamical magnetization relaxation rate $Q(T)=1/S$,~\cite{YesMalSha}
where $S$ is the total tunneling action, vanishes linearly with
temperature in the limit $T\to 0$. Indeed, the diverging contribution
$A$ to the action $S$ becomes the dominant one in the same limit. The
fact that experiments~\cite{YesMalSha} show a finite $Q(0)$ leads us
to conclude that there is some other low-energy cut-off besides
temperature. We speculate that it may appear as a consequence of a
mini-gap in the Dirac spectrum of nodal quasiparticles. Such a
mini-gap is expected in finite magnetic field as a result of curvature
terms in the dispersion around the nodal point. Its magnitude has been
estimated~\cite{Ashvin} to be $\delta=\kappa H$, $\kappa\approx 0.5
K{\rm Tesla}^{-1}$, where $H$ is the external magnetic field. Taking
the mini-gap into account, we expect the following behavior of the
magnetization relaxation rate at low temperatures:
\begin{eqnarray}
\label{Q_of_t}
Q(T) \sim \left\{
  \begin{array}{cc}
    T/\Delta, & T\gg\delta \\
    \delta/\Delta, & T\ll\delta
  \end{array}
\right.
\end{eqnarray}

\begin{figure}
\label{Q}
\includegraphics[width=200pt]{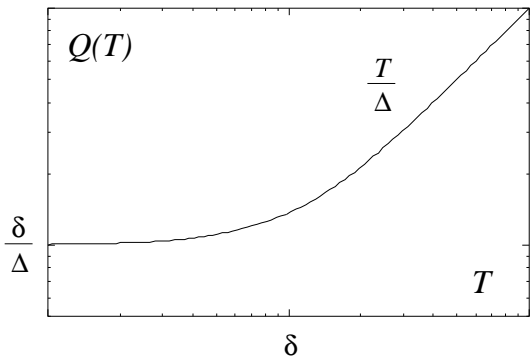}
\caption{Qualitative behavior of the dynamical magnetization
relaxation rate $Q$ as a function of temperature $T$.}
\end{figure}

This behavior is shown in Fig.~2. Qualitatively, the following picture
emerges: There is a small energy scale $\delta$ set by the size of the
mini-gap in the Dirac spectrum of nodal fermions. This energy scale
defines a cross-over temperature between two regimes: a) For
$T\gg\delta$ the function $Q(T)$ is linear in $T$ with the coefficient
$\sim 1/\Delta$ which is little dependent on the details; b) For
$T\ll\delta$ the function $Q(T)$ saturates at a value
$\sim\delta/\Delta$.  Experimentally,~\cite{YesMalSha} in the regime
$T\gg\delta$ typical slopes of $Q(T)$ are $10^{-2}K^{-1}$ which is
consistent with the typical value $\Delta\sim 10^2K$ for the Cuprates,
although the linearity in $T$ is difficult to establish. Typical
values of the cross-over temperature are $\delta\sim 1K$ (i.e. for
magnetic fields $\sim 1T$, in agreement with the theoretical
estimate~\cite{Ashvin}). The arguments above then predict $Q(0)\sim
10^{-2}$, also in agreement with experiments. Note that in this simple
picture we have omitted disorder effects on the spectrum of nodal
quasiparticles, inter-vortex interactions etc. In particular, the
Zeeman splitting is expected~\cite{Ashvin} to change the simple
$\delta=\kappa H$ behavior.

As an end note, we would like to point out one interesting consequence
of the obtained results. As we have seen, the orthogonality
catastrophe strongly inhibits the motion of half-quantum vortices,
whereas the effect disappears for doubly-quantized vortices. This
suggests a possible tendency of vortices to pair which would increase
their mobility and, effectively, decrease their energy. This scenario
is similar to the one used in recent studies of fractionalized
phases~\cite{Ring_Exchange} and deserves further investigation.

In summary, we have found that the interaction between magnetic
vortices and nodal quasiparticles in $d$-wave superconductors leads to
a strong orthogonality catastrophe in response to the motion of
half-quantum vortices. This effect strongly suppresses their mobility
and is argued to give the dominant contribution to the vortex action
at low temperatures. Based on these results, we have developed a
simple description of the quantum creep of vortices in $d$-wave
superconductors. We would like to thank A.~Paramekanti, A.~Ludwig and
M.~P.~A.~Fisher for discussions and criticism.


\end{document}